\documentclass[11pt,twoside]{article}
\usepackage{asp2010}
\resetcounters
\begin{document}

\title{Observations of transients and pulsars with LOFAR international stations}
\author{Maciej~Serylak,$^{1,2}$ Aris~Karastergiou,$^3$ Chris~Williams$^4$, Wes~Armour$^5$ and LOFAR Pulsar Working Group
\affil{$^1$ Station de Radioastronomie de Nan\c cay, Observatoire de Paris, CNRS/INSU, 18330 Nanc¸ay, France}
\affil{$^2$ Laboratoire de Physique et Chimie de l'Environnement et de l'Espace, LPC2E UMR 7328 CNRS, 45071 Orl\'eans Cedex 02, France}
\affil{$^3$ Astrophysics, University of Oxford, Denys Wilkinson Building, Keble Road, OX1 3RH}
\affil{$^4$ Oxford e-Research Centre, University of Oxford, Keble Road, OX1 3QG}
\affil{$^5$ Institute for the Future of Computing, Oxford Martin School, Oxford e-Research Centre, University of Oxford, Keble Road, OX1 3QG}
}

\begin{abstract}
The LOw FRequency ARray -- LOFAR is a new radio telescope that is moving the science of radio pulsars and transients into a new phase. Its design places emphasis on digital hardware and flexible software instead of mechanical solutions. LOFAR observes at radio frequencies between 10 and 240 MHz where radio pulsars and many transients are expected to be brightest. Radio frequency signals emitted from these objects allow us to study the intrinsic pulsar emission and phenomena such as propagation effects through the interstellar medium. The design of LOFAR allows independent use of its stations to conduct observations of known bright objects, or wide field monitoring of transient events. One such combined software/hardware solution is called the Advanced Radio Transient Event Monitor and Identification System (ARTEMIS). It is a backend for both targeted observations and real-time searches for millisecond radio transients which uses Graphical Processing Unit (GPU) technology to remove interstellar dispersion and detect millisecond radio bursts from astronomical sources in real-time using a single LOFAR station.
\end{abstract}

\section{Introduction}

The development of general-purpose GPUs, able to perform tasks done previously by Central Processing Units (CPUs) offers an attractive alternative for their application in radio-astronomy. Because the storage and off-line processing of such a vast amount of data is difficult and costly, the new generation of radio telescopes, such as LOFAR, requires High Performance Computing (HPC) solutions to process the enormous volumes of data that are typically produced during a survey for fast radio transients \citep{jwt+12}. Real-time searches for radio transients, which use GPU technology to remove interstellar dispersion and detect radio bursts from astronomical sources in real-time are now possible. Also it is important to note that real-time processing offers the possibility to react as fast as possible and to conduct follow-up observations of any event. We report here on the installation of a new backend which can be used with a single international LOFAR station.

\section{LOFAR - a brief description}

The LOFAR radio telescope designed and constructed by the Netherlands Institute for Radio Astronomy (ASTRON), is one of the largest radio telescopes ever built, using a new concept based on a vast array of omni-directional antennas grouped in 48 so-called stations. 40 of these stations are distributed in an array split between a denser core region near Exloo, the Netherlands and extending out to remote stations within the Netherlands. The remaining 8 international stations are located in the neighbouring countries: Germany, France, Sweden and the United Kingdom. Further stations can also be built and added to the array and currently a few stations are in planning stage in Germany, Poland and Ireland. Together with the Dutch stations, they make the International LOFAR Telescope (ILT). The design of the LOFAR telescope allows for multi-frequency and wide-field multi-beam observations making it suitable for imaging and spectral studies of radio astronomical sources. Details of system architecture and signal processing can be found in \citet{vgn09} and a full description of LOFAR will soon be published (van Haarlem et~al., in prep.). We limit the description only to the general information.

LOFAR observes in the frequency range of 10 to 240 MHz. To cover such a wide frequency range and to avoid the frequency modulation (FM) broadcasting bands, two types of antennas are used. First, the Low Band Antennae (LBA), can observe in the frequency range from 10 to 90 MHz, but they are optimised for a range of 30 to 90 MHz. The lower limit of 10 MHz is defined by transmission of radio waves through the Earth's ionosphere. Secondly, the High Band Antennae (HBA) cover the frequency range from 110 to 240 MHz, and consist of 16 dipoles grouped into tiles of 4 by 4 dipoles each, which are phased together using an analogue beam-former within the tile itself. A single LOFAR station can be set to observe either using LBA or HBA antennas, but not both simultaneously. However, simultaneous observations with LBA and HBA can be obtained by using multiple stations simultaneously, one using LBAs and one using HBAs.

\begin{figure}[!ht]
\centering
\includegraphics[width=\textwidth]{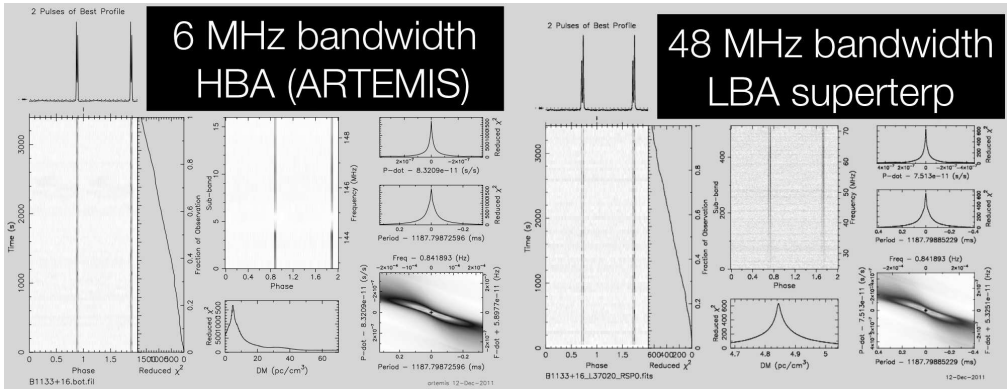}
\caption{Two diagnostic plots made from simultaneous observations done with single international LOFAR station using HBA antennas covering 6 MHz of bandwidth and recorded with the ARTEMIS backend (left) and the LOFAR core at the full observable bandwidth of 48 MHz using LBA antennas (right). One can easily notice similar signal-to-noise ratios of both observations. This example also illustrates that international LOFAR stations could be used together with the LOFAR core for simultaneous observations of sources with frequency range of 10 to 240 MHz covered by LOFAR.}
\end{figure}

The signals from individual stations are transported by a dedicated fibre-optic network to the Center for Information Technology of the University of Groningen where CEntral Processing facility (CEP) which houses a BlueGene/P supercomputer, a general purpose cluster for auxiliary processing and a long term storage archive. The main purpose of the CEP is to correlate the station data and deliver a data product which can be further processed by the user. There are many ways in which the various parts of LOFAR (antennas, tiles, stations) can be combined to perform observations. The almost completely digital nature of the LOFAR signal processing chain means that it is highly flexible to suit a particular observational goal. Pulsar observations are normally taken in a beam-formed mode, where data are recorded as a single beam pointing in the direction of the source of interest \citep{sha+11}.

\section{ARTEMIS - backend for international LOFAR stations}

The LOFAR international stations have the ability to be used independently to conduct targeted observations of known bright objects, or wide field monitoring of transient events. The international stations have greater sensitivity thanks to the doubled number of LBA antennas and HBA tiles compared to the Dutch stations. During LOFAR operations the international stations might be not used as long baselines are not currently in demand, which makes them well suited to be used as standalone instruments. As opposed to standard observations within the LOFAR network, where all the data is normally sent to, and processed by CEP, an independent backend has to be used to process the station data. This backend is called ARTEMIS: a project that is based on a versatile telescope backend using CPUs and GPUs for processing of time-series data in radio astronomy.

The ARTEMIS backend is responsible for processing the beam-formed data coming from LOFAR station. The hardware is composed of standard off-the-shelf server nodes hosting GPUs. These nodes perform (in real-time) all the operations necessary to discover short duration radio pulses from pulsars and fast transients \citep[][]{cm03, wdb+07}, thanks to a modular software structure operating in a C++ scalable framework (PELICAN, developed at the OeRC). AMPP (ARTEMIS Modular PELICAN Pipelines) is the software that was developed for receiving the data, further channelisation in finer frequency channels, generation of Stokes parameters, excision of radio frequency interference, integration, real-time dispersion searches and detection of interesting signals across multiple telescopes, in high-throughput CPU and GPU code. The detailed description of the software and algorithms used for detection can be found in Karastergiou et~al. in prep., \citet{akg+11} and \citet{mks+11}.

\end{document}